\documentclass[pre,aps,twocolumn,showpacs]{revtex4}
\usepackage{epsfig}
\usepackage{graphicx}
\begin{document}
\title{The Knudsen temperature jump and the Navier-Stokes hydrodynamics of granular gases driven by thermal walls}
\author{Evgeniy Khain$^{1}$, Baruch Meerson$^{2}$, and Pavel V. Sasorov$^{3}$}

\affiliation{$^{1}$Department of Physics, Oakland University,
Rochester MI 48309, USA}

\affiliation{$^{2}$Racah Institute of Physics, Hebrew University
of Jerusalem, Jerusalem 91904, Israel}

\affiliation{$^{3}$Institute of Theoretical and Experimental
Physics, Moscow 117218, Russia}

\begin{abstract}
Thermal wall is a convenient idealization of a rapidly vibrating
plate used for vibrofluidization of granular materials. The objective of
this work is to
incorporate the Knudsen temperature jump at thermal wall
in the Navier-Stokes hydrodynamic modeling
of dilute granular gases of monodisperse particles that
collide nearly elastically. The Knudsen temperature jump manifests
itself as an additional term, proportional to the temperature gradient, in the boundary condition for the
temperature. Up to a numerical pre-factor ${\cal O}(1)$, this term
is known from kinetic theory of elastic gases. We determine the previously unknown
numerical pre-factor by measuring, in a series of molecular dynamics
(MD) simulations, steady-state temperature profiles of a gas of
elastically colliding hard disks, confined between two
thermal walls kept at different temperatures, and comparing the
results with the predictions of a hydrodynamic calculation employing
the modified boundary condition. The modified boundary condition is
then applied, without any adjustable parameters, to a hydrodynamic
calculation of the temperature profile of a gas of
inelastic hard disks driven by a thermal wall. We find the
hydrodynamic prediction to be in very good agreement with MD
simulations of the same system. The results of this work pave the
way to a more accurate hydrodynamic modeling of driven granular
gases.
\end{abstract}
\pacs{47.70.Nd, 45.70.-n, 51.10.+y} \maketitle

\section{Introduction}

Granular gas - a shorthand for a rapid flow of a low-density
assembly of inelastically colliding particles - continues to attract
much attention \cite{gasreviews,hydrreview,brilliantov}.
Not in the least this is because  of fascinating pattern-forming
instabilities that develop in granular gases: clustering,
convection, phase separation, oscillatory instability \textit{etc.},
see \textit{e.g.} Ref. \cite{Aranson} for a review.  A continuum
description of these phenomena is provided by the Navier-Stokes
granular hydrodynamics which is derivable, under certain
assumptions, from the more basic kinetic theory. For dilute gases
this is the Boltzmann equation, properly modified to account for
inelastic particle collisions. Within the Chapman-Enskog gradient
expansion formalism the above-mentioned assumptions are not specific to
granular gases and are the same as in hydrodynamics of
\textit{elastic} hard sphere fluid: (i) the mean free path (and the
mean time between two consecutive particle collisions) of the gas should be
much smaller than any length scale (correspondingly, time scale)
that one attempts to describe hydrodynamically, and (ii) the gas
density should be much smaller than the close-packing density of spheres.
It is crucial that the validity of these assumptions cannot be guaranteed
\textit{a priori}, as they operate with quantities that become
explicitly known only after the hydrodynamic problem in question is
solved. Furthermore, as it has been found in numerous recent
examples, inelasticity of particle collisions drives strong
gradients of hydrodynamic fields. As a result, the scale separation
condition [condition (i) above] usually breaks down unless the
particle collisions are \textit{nearly elastic}. The nearly elastic
limit is quite restrictive, as it puts a vast majority of granular
materials beyond the formal limits of the Navier-Stokes
hydrodynamics. Still, this limit proved to be very useful because, with
its great predictive power and readily available imagery of
macroscopic flow patterns, the Navier-Stokes hydrodynamics gives a valuable
insight into complex collective phenomena in granular flows that, at
least qualitatively, often persist well beyond the nearly elastic limit.

A direct quantitative measure of scale separation in a  gas (both
molecular, and granular) is the Knudsen number $K$: the ratio
of the (local) mean free path of the gas to a characteristic
hydrodynamic length scale. Starting from the Boltzmann equation one
obtains, in the zero order approximation in $K \ll 1$, \textit{ideal}
hydrodynamics: the Euler hydrodynamics for the gas of elastically colliding
spheres, and the Euler hydrodynamics with bulk energy losses for the
granular gas \cite{brilliantov,fouxon1,fouxon2}. In the next, first
order in $K$ one obtains the Navier-Stokes hydrodynamics
\cite{Chapman,LP,brilliantov}. The still higher, second order in $K$ brings two
types of new effects: the Burnett correction terms in the hydrodynamic
equations and ${\cal O}(K)$ corrections to the boundary conditions
\cite{Chapman,LP}. The Knudsen numbers of granular flows are
typically not very small, therefore the second order effects are often
important.  It is crucial that, in many cases of interest, the
corrections to the boundary conditions are more important than the
Burnett correction terms in the hydrodynamic equations. (See,
\textit{e.g.} Ref. \cite{LP} for a detailed explanation of this fact
in the case of molecular gases. This explanation typically holds for
dilute granular gases with nearly elastic particle collisions.) The
present work deals with a quantitative incorporation of one of the
corrections to the boundary conditions - the one corresponding to
the Knudsen temperature jump at a thermal wall - in the
Navier-Stokes hydrodynamic modeling of dilute granular gases of
nearly elastically colliding particles.

The Knudsen temperature jump, and other types of Knudsen jumps/slips
of hydrodynamic fields at the system boundaries \cite{Chapman,LP}, are
intimately related to what is called the Knudsen layer: a
next-to-wall region which thickness is comparable to the (local)
mean free path of the gas, and where therefore hydrodynamic theory breaks
down.  Although well known in the context of rarefied molecular
gases \cite{Chapman,LP,Kogan}, the physics of the Knudsen layer and
its consequences for the bulk flow have received only a cursory
attention from the granular community \cite{GoldhirschChaos}. This
is in spite of the fact that the Knudsen numbers of granular flows
are typically not small, and temperature jumps were evidently
present in, and noticed by the authors of,  a number of MD
simulations of granular gases driven by thermal walls, see
\textit{e.g.} Refs. \cite{Grossman,Ramirez,Soto}. Furthermore, it
was observed \cite{Grossman} that, inside the Knudsen layer, the
particle velocity distribution strongly deviates from a
Maxwellian, as the temperature of the particles moving toward the
thermal wall is different from that of the outgoing particles.

To our knowledge, the first detailed quantitative study of the role
of Knudsen layers in granular gases is the recent work by Galvin
\textit{et al.} \cite{Hrenya}. They  performed three-dimensional MD
simulations of a monodisperse granular gas driven by two opposite
thermal walls, and compared steady-state hydrodynamic fields, and
the steady-state heat flux through the system, with predictions from
the Navier-Stokes hydrodynamics with two different sets of
constitutive relations accounting for finite-density corrections in
the spirit of the Enskog theory. Galvin \textit{et al.} did not
attempt to modify the boundary conditions at the thermal walls. By
measuring the deviations between the hydrodynamic theory and
simulations, they estimated the effective Knudsen layer thickness as
$2.5$ (local) mean free paths. They also observed that Navier-Stokes
hydrodynamic calculations that do not account for the presence of
Knudsen layers remain accurate ``for Knudsen layers collectively
composing up to 20\% of the domain" \cite{Hrenya}.

Essentially, the objective of Galvin \textit{et al.} \cite{Hrenya}
was to establish the validity limits of the Navier-Stokes
hydrodynamics that \textit{ignores} the presence of the Knudsen
layers and the corrections to the boundary conditions that appear in
the second-order expansion in $K$. The objective of the present work
is different: we will take these corrections into account for the
purpose of a more accurate description of the hydrodynamic fields in
the bulk  - outside the Knudsen layers.

The strategy that we suggest makes
the full use of a crucial simplification that appears in the limit of
nearly elastic particle collisions: the only limit we will focus on. Here
the correction term in the boundary condition for the temperature is
independent, in the leading order of the theory, of the particle
collision inelasticity and can therefore be adopted from the theory
of elastically colliding hard sphere fluid.  The corrected boundary condition must
be satisfied by the hydrodynamic fields \textit{extrapolated} to the
boundary from the bulk, rather than by the true local values of the
fields at the boundary; see Ref. \cite{LP} for a pedagogical review
of this important circumstance. Furthermore, up to a single
numerical pre-factor, this term is known from kinetic theory
\cite{Chapman,LP}.  We will \textit{not} attempt to find this pre-factor by
solving the Boltzmann equation inside the Knudsen layer (such a
solution would be quite involved) and matching the ``inner" kinetic solution
next to the boundary with the ``outer" hydrodynamic solution in the
bulk. Instead, we will extrapolate the steady-state temperature
profiles in the bulk, measured in MD simulations, to the boundaries.
As will be shown shortly, a comparison of the extrapolated values of
the gas temperature with those predicted from the hydrodynamics
which employs the modified boundary condition yields an estimate of
the unknown numerical pre-factor: the only adjustable parameter of
our theory. Once the pre-factor is found, the modified boundary
condition (with no adjustable parameters) can be used for, and render
a more accurate hydrodynamic description of, a host of
two-dimensional gases (of either elastic, or weakly inelastic
particles, with and without gravity) driven by thermal walls of the
same type.

The remainder of the paper is organized as follows. We start Section
II with Navier-Stokes hydrodynamic calculations of a steady-state
temperature profile of a two-dimensional gas of monodisperse {\it
elastic} hard disks, confined between two thermal walls kept at
different temperatures. The same Section II reports a series of MD
simulations of the same system. A comparison between the two yields
the unknown numerical pre-factor of the correction term
of the modified boundary condition. The
modified boundary condition is then
applied in Section III to a gas of {\it inelastic} hard disks
driven by a single thermal wall, and the results are compared with
those of MD simulations of the same system. Section IV presents a
brief discussion of our results and of related open
questions.

\section{Knudsen temperature jump and modified boundary conditions}

Consider a dilute assembly of elastically colliding monodisperse
hard disks in a two-dimensional box, confined between two thermal
walls located at $x=-L_x/2$ and $x=L_x/2$ and kept at different
temperatures $T_1$ and $T_2<T_1$, respectively. What is the
steady-state temperature of the gas next to the thermal wall
$x=-L_x/2$? There are two different groups of particles here: the
outgoing particles with temperature $T_1$ and the incoming particles
with a smaller temperature. Therefore, the overall gas temperature
(defined as the average energy of particles) next to the wall
$x=-L_x/2$ is {\it smaller} than $T_1$. By the same argument, the
gas temperature next to the wall $x=L_x/2$ is {\it larger} than
$T_2$. This is the well known Knudsen temperature jump effect. For
small Knudsen numbers, $K \ll 1$, the corrected boundary conditions
that accommodate the Knudsen temperature jump \textit{for the
purpose of hydrodynamic calculations in the bulk} have the following
form \cite{Chapman,LP}
\begin{eqnarray}
&&\left(T-g_0\,\lambda\,\frac{\partial T}{\partial x}\right)\Big{|}_{x=-L_x/2}\, =
T_1\,,
\nonumber \\
&&\left(T+g_0\,\lambda\,\frac{\partial T}{\partial x}\right)\Big{|}_{x=L_x/2}\, =
T_2\,, \label{bound1}
\end{eqnarray}
where $\lambda = (2 \sqrt{2}\,d\,n)^{-1}$ is the local
mean free path, $d$ is the particle diameter, $n$ is the number
density of the gas, and $g_0={\cal O}(1)$ is a numerical pre-factor that depends on the
exact nature of the boundary. In the
following we will assume a most commonly used thermal wall protocol and
determine the unknown pre-factor from MD simulations. As $d T/d x < 0$,  conditions (\ref{bound1}) imply that $T <
T_1$ near the wall $x=-L_x/2$, and $T > T_2$ near the wall
$x=L_x/2$, as expected.

Now we employ the Navier-Stokes hydrodynamic (or rather \textit{hydrostatic})
equations to describe the steady state of this system with a zero
mean flow:
\begin{equation}
p=const\;\;\; \mbox{and} \;\;\;\mathbf{\nabla}\cdot(\kappa
\mathbf{\nabla} T)=0\,, \label{E12}
\end{equation}
where $p$ is the gas pressure, and $\kappa$ is the thermal
conductivity. To make the formulation closed we need to specify
the equation of state $p=p(n,T)$ and an expression for
$\kappa$ in terms of $n$ and $T$. For a dilute gas of elastically colliding
hard disks of
mass $m=1$ and diameter $d$ these are given by the well known relations
\begin{equation}
p=n\,T \;\;\; \mbox{and} \;\;\;
\kappa=\kappa_0\,\frac{2\,T^{1/2}}{\pi^{1/2}\,d}\,, \label{const1}
\end{equation}
where the pre-factor $\kappa_0=1.029$ appears \cite{Gass} in the third
Sonine polynomial approximation \cite{Burnett,Liboff}. The normalization condition
\begin{equation}
\int_{-L_x/2}^{L_x/2} dx \int_{-L_y/2}^{L_y/2} dy\, n(x,y)\,=\,N\,,
\label{norm1}
\end{equation}
fixes the total number of particles $N$. We are looking for
a $y$-independent solution, $n=n(x)\,,T=T(x)$, and
rescale variables: ${\mathbf{r}}/L_x \to
\mathbf{r}\,,T/T_2 \to T\,,n/\bar{n} \to n$, where
$\bar{n}= N/ (L_x L_y)$ is the average number density of the gas.
Equations (\ref{E12}) become
\begin{equation}
P=const\;\;\; \mbox{and} \;\;\; \frac{d}{d x}\left(T^{1/2} \frac{d
T}{d x}\right)=0\,, \label{E2}
\end{equation}
where $P = p/(\bar{n}\,T_2)$ is the rescaled
pressure. The boundary conditions (\ref{bound1}) become
\begin{eqnarray}
&&\left(T-\frac{g_0\,K}{n}\,\frac{d T}{d x}\right)\Big{|}_{x=-1/2}\, =
\delta\,,
\nonumber \\
&&\left(T+\frac{g_0\,K}{n}\,\frac{d T}{d x}\right)\Big{|}_{x=1/2}\, =
1\,, \label{bound2}
\end{eqnarray}
where $K = L_y/(2\sqrt{2}\,N\,d) \ll 1$ is the effective Knudsen number of the system, and
$\delta = T_1/T_2$. The normalization condition
(\ref{norm1}) becomes
\begin{equation}
\int_{-1/2}^{1/2} n(x)\,dx\,=\,1\,. \label{norm2}
\end{equation}
Solving the second of Eq.~(\ref{E2}), we arrive at the temperature profile
\begin{equation}
T(x) = (A\,x\,+\,B)^{2/3}\,. \label{solution}
\end{equation}
Now we treat the ${\cal O}(K)$ terms in Eqs.~(\ref{bound2}) as small
corrections and obtain the constants $A$ and $B$ up to the first
order in $K$:
\begin{eqnarray}
&&A=1-\delta^{3/2}+3\,K\,g_0\,(\delta+1)(\delta^{1/2}-1)\,,
\nonumber \\
&&B=\frac{1+\delta^{3/2}}{2}-\frac{3}{2}\,K\,g_0\,(\delta-1)(\delta^{1/2}-1)\,.
\label{AB}
\end{eqnarray}
The predicted effective temperature jumps,
\begin{equation}
\Delta T_{-}  =  2\,K\,g_0\,\delta^{1/2}\,(\delta^{1/2}-1)
\label{deltaT1}
\end{equation}
at the wall $x=-1/2$, and
\begin{equation}
\Delta T_+  =  2\,K\,g_0\,(\delta^{1/2}-1) \label{deltaT2}
\end{equation}
at the wall $x=1/2$, are proportional to the yet unknown numerical pre-factor $g_0$.
\begin{figure}[ht]
\centerline{\includegraphics[width=9.0cm,clip=]{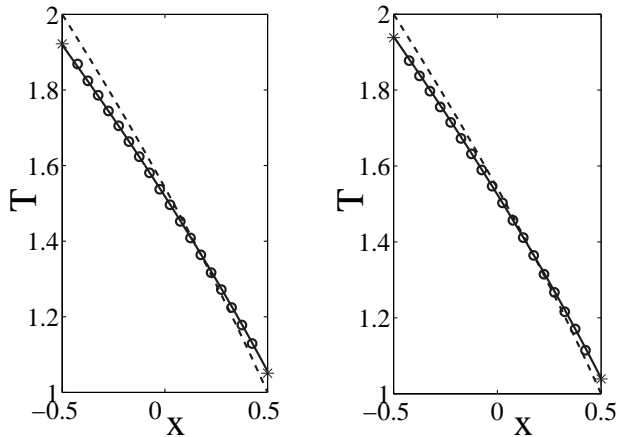}}
\vspace{0.2cm} \caption{Steady-state temperature versus the rescaled coordinate $x$ for a
dilute gas of elastically colliding disks, confined by two thermal
walls for $\delta = T_1/T_2 = 2$. The circles: MD simulations. The
solid lines: predictions from the Navier-Stokes hydrodynamics with the
modified boundary conditions  Eqs.~(\ref{solution}) and
(\ref{AB}) with $g_0=5.0$.  The dashed lines: predictions from
the Navier-Stokes hydrodynamics with the boundary conditions
 $T(x=-1/2)=\delta$
and $T(x=+1/2)=1$ \textit{not} including the ${\cal O}(K)$ terms.
The parameters are $\bar{n} = 0.04$, $K=0.01414$,
$N=2000$, $L_x = 625$ and $L_y = 80$ (the left panel) and $\bar{n} =
0.01$, $K=0.01$, $N=16284$, $L_x = 3540$ and $L_y = 460$ (the right
panel). The wall temperatures, indicated by the asterisks, were
determined by linear extrapolation of $T^{3/2}(x)$ from the bulk to
the corresponding wall.} \label{temperature}
\end{figure}

To test the hydrodynamic predictions and find the unknown numerical
pre-factor $g_0$, we performed a series of  MD simulations, using an
event driven algorithm described in Ref. \cite{Rapaport}. The walls
at $y=\pm L_y/2$ were assumed elastic. The
thermal walls at $x=\pm L_x/2$ were implemented in the simulations in the following
way: upon a collision with a thermal wall the normal component of
the particle velocity is drawn from a Maxwell distribution with the
prescribed wall temperature, while the tangential component of the
particle velocity remains unchanged. For each set of parameters we
started the simulation from an initially uniform spatial particle
distribution and a Maxwell velocity distribution with temperature
$T_2$, and waited until the gas reached a steady state. This was
verified by analyzing the time dependence of the $x$-component of
the center of mass of the gas, and the time dependence of the
temperature next to the thermal wall at $x=-L_x/2$. Then we computed
the steady-state temperature profile of the gas by averaging
instantaneous temperature profiles over a long time. Figure
\ref{temperature} shows two of the many steady-state temperature
profiles measured in our MD simulations. The effective temperature
at each of the two walls was obtained by linear extrapolation of the
profile of $T^{3/2}(x)$ from the bulk to the wall.

\begin{figure}[ht]
\centerline{\includegraphics[width=7.2cm,clip=]{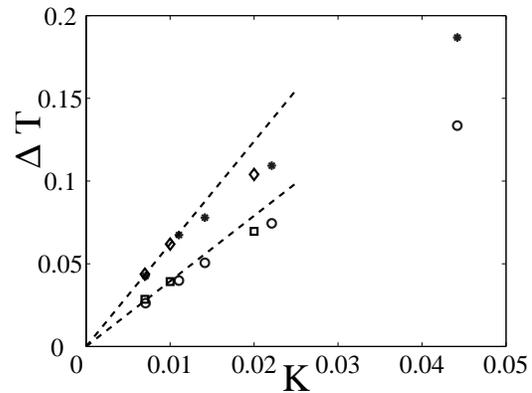}}
\caption{Effective temperature jumps at each of the two thermal
walls versus the Knudsen number $K$, as measured in MD simulations
(symbols) and predicted by Eqs.~(\ref{deltaT1})-(\ref{deltaT2})
(dashed lines). The lower dashed line, circles and squares
correspond to the right wall: $x = 1/2$ in the
rescaled units. The upper dashed line,
diamonds and asterisks correspond to the left wall, $x = -1/2$ in the rescaled units. MD
simulations were performed for two different average densities:
$\bar{n} = 0.04$ (circles and asterisks) and $\bar{n} = 0.01$
(squares and diamonds). At $K\alt 0.02$ a linear dependence of the
effective temperature jumps on $K$ is observed. The linear-dependence part yields $g_0= 5.0 \pm 0.3$.  In the simulations with
$\bar{n} = 0.04$ we varied $K$ by varying the system dimensions
$L_x$ and $L_y$ and keeping a constant total number of particles
$N=4000$. In the simulations with $\bar{n} = 0.01$, $L_y=460$ was kept
fixed, while $L_x$ and $N$ were varied from $L_x=1770$ and $N=8142$,
for largest $K$, to $L_x=5050$, $N=23230$ for smallest $K$. In all
simulations we kept $\delta = T_1/T_2 = 2$.} \label{deltaT}
\end{figure}

Figure~\ref{deltaT} shows the effective temperature jump at each of
the two thermal walls as found in our MD simulations at different
but small Knudsen numbers and different but small average gas
densities. The results for different densities practically coincide.
The same figure shows the hydrodynamic predictions from
Eqs.~(\ref{deltaT1})-(\ref{deltaT2}). One can see that a linear
dependence of the effective temperature jumps on $K$ is observed
only for quite small Knudsen numbers: $K\alt 0.02$.  This linear
dependence part yields the pre-factor $g_0$:
\begin{equation}
g_0 = 5.0 \pm 0.3\,. \label{constant}
\end{equation}
Examples of the resulting hydrodynamic temperature profiles, see
Eqs.~(\ref{solution}) and (\ref{AB}), are depicted in  Figure
\ref{temperature}, and very good agreement with the MD simulations
is observed. The same figure also shows the hydrodynamic temperature
profiles calculated with the boundary conditions  $T(x=-1/2)=\delta$ and
$T(x=+1/2)=1$ \textit{not}
including the ${\cal O}(K)$ terms. As expected, they show a worse agreement with
the MD simulations.

In the next section we apply the modified boundary condition with the same value of $g_0$
to a \textit{granular} gas driven by a thermal wall of the same type.

\section{Granular gas driven by a thermal wall}

Having determined the pre-factor $g_0$, we can now use the modified
boundary condition with this $g_0$ for a more accurate hydrodynamic description of
two-dimensional gases (of either elastic, or weakly inelastic
particles, with and without gravity) driven by thermal wall(s) of
the same type. As an example, we will consider here a simple, indeed
prototypical, setting: a granular gas confined in a two-dimensional
box and driven by a single thermal wall at zero gravity. In a steady
state, the energy supplied into this gas from the thermal wall is
balanced by the collisional energy loss. The simplest steady state
of this system is the so called ``stripe state", where the gas
density and temperature fields do not depend on the coordinate
parallel to the thermal wall. We will consider the region of
parameter where this simple steady state is hydrodynamically stable,
see Refs. \cite{LMS,KM} for detail.

Let us complete the specification of the model. We consider a dilute assembly of
inelastically colliding hard disks of diameter $d$ and mass $m=1$,
moving in a box with dimensions $L_x \times L_y$ at zero gravity.
Collisions of disks with the walls $x=0$ and $y=\pm L_y/2$ are
assumed elastic. The thermal wall, kept at $T=T_0$, is located at $x = L_x$. The inelasticity
of the particle collisions
is parameterized by a constant coefficient of normal restitution  $r$;
we assume the nearly elastic limit $1-r^2 \ll 1$.

Due to the inelastic particle collisions, the
granular temperature decreases with an increase of the distance
from the thermal wall.  To
maintain a constant pressure in a steady state, the particle density must
increase with this distance, reaching its maximum value next to the opposite
(elastic) wall. The steady state density and temperature profiles of the stripe state
are described by
the following hydrodynamic/hydrostatic equations [compare with Eqs. (\ref{E2})]:
\begin{equation}
p=const\;\;\; \mbox{and} \;\;\;\frac{d}{dx}\left(\kappa \frac{d T}{dx}\right)=I\,.
\label{E52}
\end{equation}
For the nearly elastic collisions, the granular pressure $p$ and the thermal conductivity
$\kappa$ are still given by Eq.~(\ref{const1}), while
\begin{equation}
I= \sqrt{\pi} \,(1-r^2) d n^2 \,T^{3/2} \simeq  2\sqrt{\pi} \,(1-r) d n^2\, T^{3/2} \label{loss}
\end{equation}
is the energy loss rate of the gas due to the particle collisions, see\textit{ e.g.} Ref. \cite{brilliantov},
in the limit of $1-r\ll 1$.

Let us rescale the $x$-coordinate by $L_x$, the number density of the gas by $\bar{n}$, and the gas
temperature by $T_0$. Introducing
a rescaled inverse density $z(x)=\bar{n}/n(x)$ and
rescaled pressure $P=p/(\bar{n} T_0)$, we can rewrite the
second equation in Eqs.~(\ref{E52}) in the following form
\begin{equation}
(z^{3/2})^{\prime\prime}=3\, \xi^2\,z^{-1/2}, \label{E62}
\end{equation}
where the primes stand for the derivatives with respect to the rescaled coordinate $x$,
\begin{equation}
\xi=\sqrt{\frac{\pi (1-r)}{2 \kappa_0}}\,\bar{n} \,d\, L_x
\end{equation}
is a hydrodynamic inelastic loss parameter, and we recall that $\kappa_0=1.029$.
The inelastic loss parameter $\xi$ defines the characteristic hydrodynamic length scale
of the problem $\xi^{-1}$ or, in the dimensional variables, $\xi^{-1} L_x$.

One boundary condition for Eq.~(\ref{E62}) can be specified at the elastic wall:
\begin{equation}
z^{\prime}(x=0)=0\,. \label{bound40}
\end{equation}
Conservation of the total number of particles yields the normalization condition
\begin{equation}
\int_{0}^{1} \frac{d x}{z(x)} = 1\,. \label{conservation}
\end{equation}
Finally, the modified boundary condition at the thermal wall $x=1$
is given by
\begin{equation}
P z \left(1+ g_0 K \frac{d z}{d x}\right)\Big{|}_{x=1} = 1\,.
\label{bound4}
\end{equation}
There are two unknowns in this formulation: the rescaled inverse
density $z(x)$ and the constant rescaled pressure $P$. The density profile is completely determined by Eq.~(\ref{E62}) and
conditions (\ref{bound40}) and (\ref{conservation}), and can be found
analytically in terms of $x=x(z)$ \cite{KM}. The result is:
\begin{equation}
x=\frac{z_0}{2\xi}\left[\mbox{arccosh}\sqrt{\frac{z}{z_0}}+\sqrt{\left(\frac{z}{z_0}\right)^2-\frac{z}{z_0}}\right]\,,
\label{analytic}
\end{equation}
where $$z_0=\frac{2\xi}{\xi+\sinh(\xi)\cosh(\xi)}\,.$$ Note
that $z(x=1)=z_0 \cosh^2(\xi)$.

In contrast to the density profile, which is independent
of the boundary condition (\ref{bound4}), the temperature profile is sensitive to
the ${\cal O}(K)$ correction in Eq.~(\ref{bound4}).
Using Eq.~(\ref{bound4}), we calculate the pressure up to the first order in
$K$, so that the resulting temperature profile $T(x)=Pz(x)$ is
\begin{equation}
T(x)=\frac{1-2 g_0 K \xi \tanh(\xi)}{z_0\cosh^2(\xi)} z(x).
\label{temp}
\end{equation}

\begin{figure}[ht]
\centerline{\includegraphics[width=7.0cm,clip=]{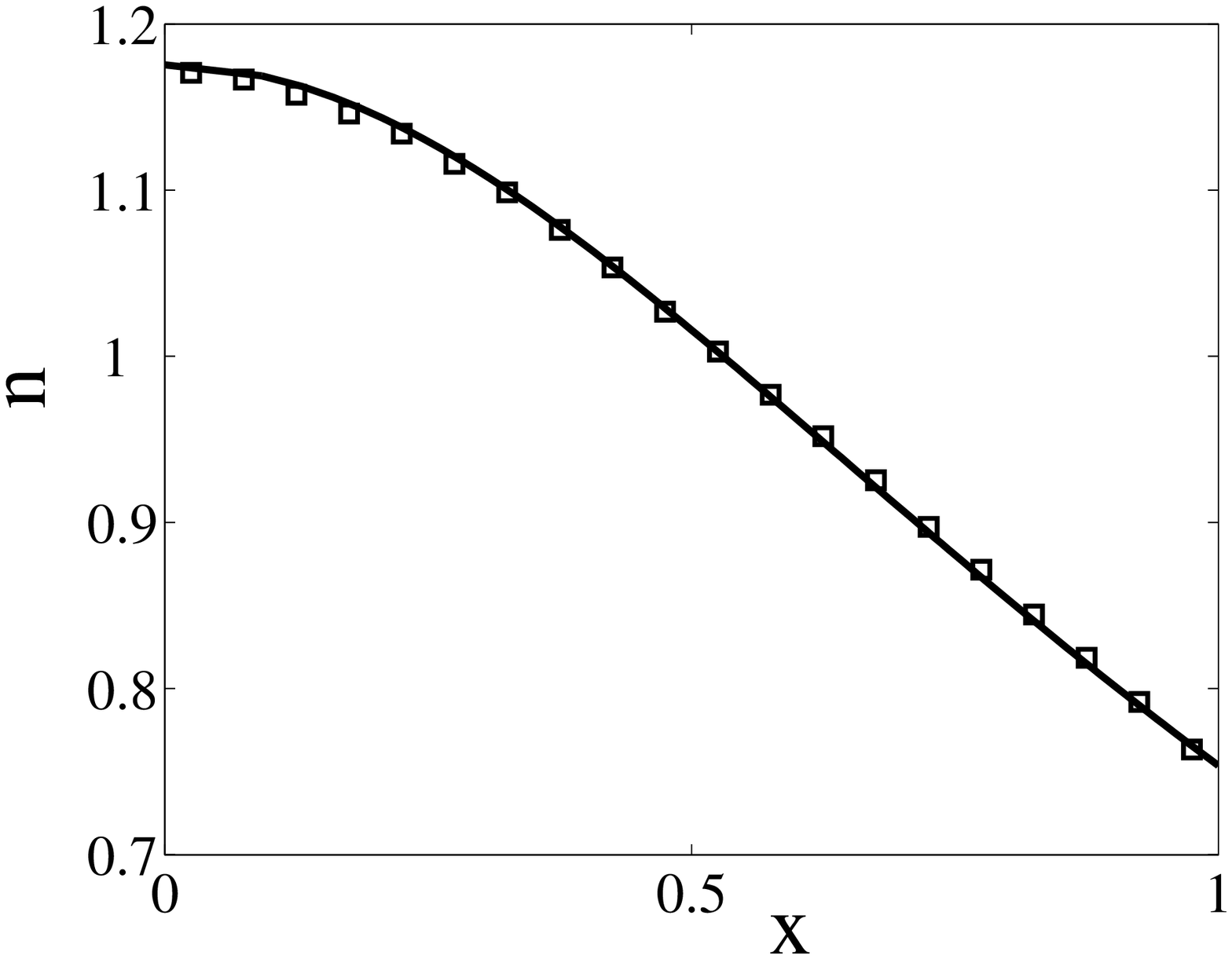}}
\centerline{\includegraphics[width=7.0cm,clip=]{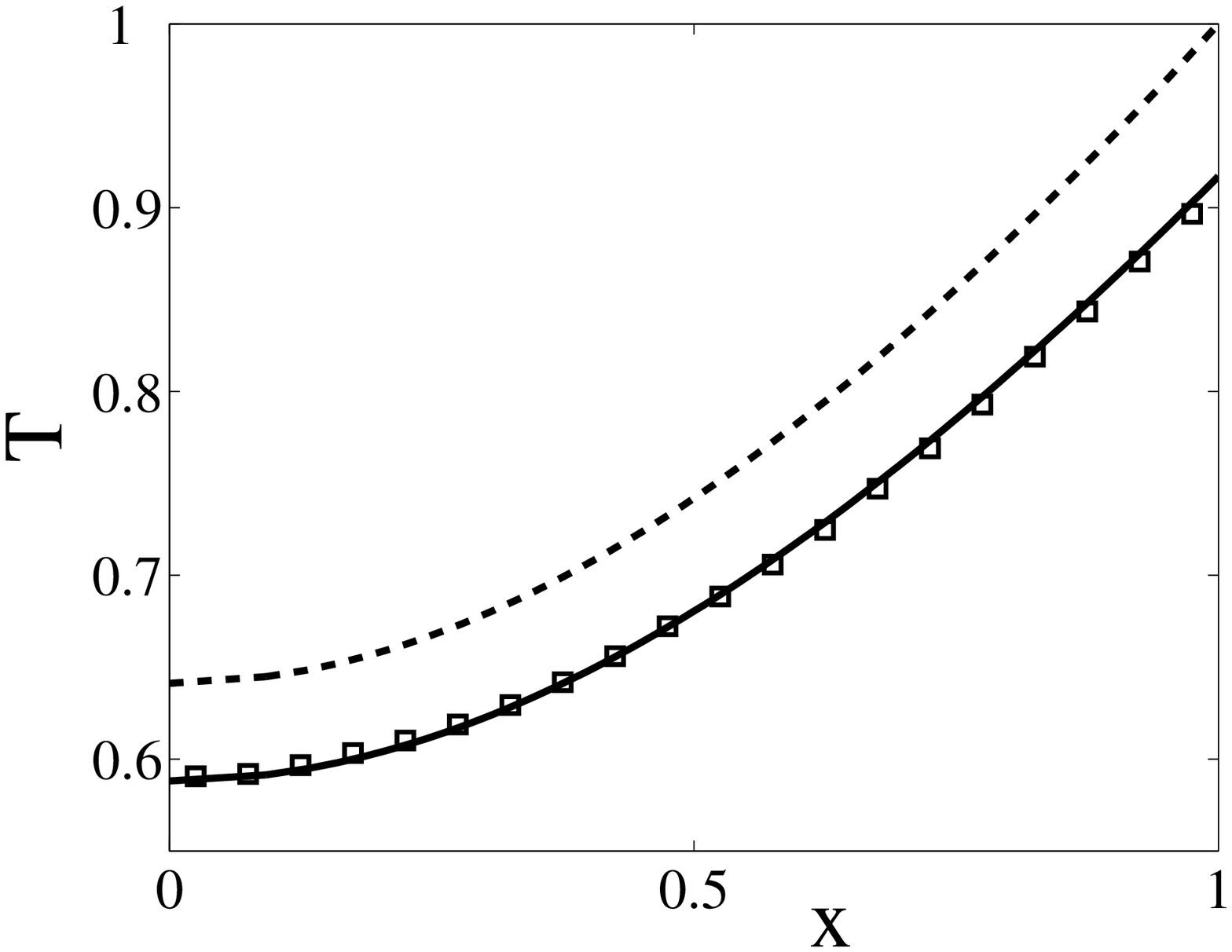}}
\caption{Density (the upper panel) and temperature (the lower panel)
profiles of a gas of inelastic hard disks driven by a
thermal wall. The theoretical profiles (solid curves) are in
good agreement with MD simulations (squares). Shown by the dashed
line is the
temperature profile obtained without the Knudsen correction.
The parameters $N = 3540$, $L_x=1770$, $L_y=200$, and
$r = 0.999$ correspond to $\xi
= 0.6916$, $K=0.02$, and $\bar{n}=0.01$.} \label{stripe}
\end{figure}

The hydrodynamic density profile $n(x) = 1/z(x)$, given by
Eq.~(\ref{analytic}), and temperature profile,
given by Eq.~(\ref{temp}), are shown in Fig.~\ref{stripe} (the solid lines in the upper
and lower panels, respectively). To test the hydrodynamic
results, we performed event-driven MD simulations \cite{Rapaport}
and measured the density and temperature in this system. As before,
we verified that the system was in the steady state and computed the
profiles, performing averaging over a long time. Figure \ref{stripe}
shows (by the squares) the density (the upper panel) and temperature (the
lower panel) as observed in MD simulations. In both
cases there is a good agreement between the theoretical profiles and
those observed in MD simulations. For comparison, the dashed
line in the lower panel of Fig. \ref{stripe} shows the
hydrodynamic temperature profile obtained when \textit{ignoring} the Knudsen
correction in the boundary condition (\ref{bound4}). The observed
disagreement with
the results of MD simulations clearly shows
the importance of the Knudsen correction in this example.

Once the hydrodynamic problem is solved, we should verify the scale separation (and the validity of the Navier-Stokes hydrodynamics) by demanding
that the mean free path of the gas $\sim (d\bar{n})^{-1}$ be much smaller than the hydrodynamic length scale $\sim  L_x/\xi \sim (\sqrt{1-r}\,d\bar{n})^{-1}$. This yields a restrictive condition $\sqrt{1-r}\ll 1$ (first obtained in Ref. \cite{Grossman}), thus justifying \textit{a posteriori} our focus on the nearly elastic case.

\section{Summary and discussion}

We have incorporated  the Knudsen temperature jump at thermal
wall in the Navier-Stokes hydrodynamic description of
weakly inelastic dilute gases of smooth hard disks. We have
shown that this procedure may considerably improve the accuracy of
hydrodynamic calculations. Therefore, the results of this work pave the way to
a more accurate hydrodynamic modeling of driven granular gases. This is important in view of the continuing tests (and the ongoing debate on the validity
range) of the Navier-Stokes granular hydrodynamics as a quantitatively accurate theory.

In the prototypical example of a granular gas heated by
a thermal wall at zero gravity, that we have considered in this work, the modification of the boundary condition affects only the gas temperature in the bulk, and does not affect the density profile. In more general settings (such as those including gravity), the density profile will be affected as well.

Future work should attempt to extend our approach to other types of
Knudsen jumps/slips at the boundaries of rapid granular flows, again
in analogy to what has been done in this context for molecular fluids
\cite{Chapman,LP}. Future work also needs to go beyond the dilute limit and
account for
finite-density corrections. A practical approach here would be to
use the Carnahan-Starling equation of state \cite{Carnahan} and
Enskog-type transport coefficients \cite{Jenkins}, but still assume
nearly elastic collisions.  The finite-density case in two dimensions may present difficulties
because of the long-lived large-scale hydrodynamic
fluctuations
which contribute to the transport in addition to the ``usual" gradient contributions \cite{cohen}.
These additional contributions formally appear as divergences of the
transport coefficients with the system size. Indeed,
we observed a clear signature of the apparent divergence of the heat conductivity in our MD simulations when
attempting to extend the Knudsen correction to the boundary condition to a moderately dense granular gas in two dimensions. In three dimensions, however, such an extension looks
promising.

\vspace{0.5cm}
\begin{acknowledgments}
We are grateful to J.R. Dorfman and M.H. Ernst for an illuminating discussion of the  transport
properties of finite-density gases of hard spheres in two dimensions.
This work was supported by the Israel Science
Foundation (grant No. 107/05), by the German-Israel Foundation
for Scientific Research and Development (Grant I-795-166.10/2003),
and by the Russian
Foundation for Basic Research (Grant No. 05-01-000964).
\end{acknowledgments}

\end{document}